# Radio-far-infrared Correlation and Cosmic Ray Electrons in Starburst Galaxies and Ultra Luminous Infrared Galaxies

T. N. Rengarajan.
*206, Vigyan, Sector 17, Navi Mumbai 400703, India*

The ubiquitous and tight correlation between the radio continuum emission and the far-infrared emission is also found to hold true for starburst galaxies and Ultra Luminous Infrared Galaxies wherein most of the emission is from compact structures. The infrared photon density in the cores is high enough for the inverse Compton scattering to dominate over the synchrotron emission thus spoiling the correlation. The high magnetic field postulated to avoid this, also encounters the problem of steep radio index that is not observed. Further, at high densities the flux of secondary electrons can exceed that of the primary. The problems posed by such effects are discussed and suggestion for further work made.

## 1. Introduction

In recent years, one of the most studied correlation in astronomy has been the ubiquitous and surprisingly tight correlation between the seemingly unrelated far infrared (FIR) and radio continuum (RC) emissions of galaxies [1-3]. The FIR emission of galaxies mainly results from thermal emission from dust which is heated by stellar radiation while the RC emission is mainly non-thermal (NT) synchrotron radiation by high energy electrons in the interstellar medium. What was surprising was that the correlation was seen over a large range of luminosities in a variety of galaxies at different radio frequencies and the dispersion of the ratio of luminosities was less than a factor of 2 [4-11]. The general basis of the correlation is the common origin of massive stars as the progenitors- the FIR reflecting the OB star luminosity and the non-thermal RC emission being related to supernovae (SN). However, there are many parameters associated with the production of these two radiations. Hence the question why the correlation is so tight has been the target of several studies. When IRAS satellite revealed the existence of Ultra Luminous Infrared Galaxies (ULIRG) in which the FIR emission is in excess of $10^{11.5}$ $L_O$ and the radio emission is from compact cores of size 100-200 pc, it was noted with surprise that the radio-FIR correlation for these galaxies follows the same trend as for the lower luminosity galaxies. In this paper I discuss the effect of high intensity and presumably high density of such cores on the radio spectral index as well as the role of secondary electrons.

## 2. Radio-FIR Correlation in Ultra Luminous Infrared Galaxies

A major discovery of the IRAS mission was Ultra Luminous Infrared Galaxies which have FIR luminosities $> 10^{11.5}$ $L_O$ and have high FIR to blue ratio. These galaxies are powered by an intense burst of star formation triggered by interaction or close encounter with another galaxy and have large concentration of gas and dust in highly compact cores. A remarkable feature is that these galaxies also follow the general trend of the radio-FIR correlation. Condon and Broderick et al [8] studied a sample of 40 ULIRGs chosen from the IRAS Bright Galaxy Catalog and obtained high resolution maps using the VLA. Most of the observed galaxies harbour compact sub-arc second 8.44 GHz sources. In many of these, the compact source of size 100-200 pc contains the bulk of the emission of the galaxy. In the radio-FIR plot, these galaxies occupy the region expected on the basis of extrapolation from the lower luminosity galaxies and the value of q, the logarithmic ratio of FIR and RC fluxes defined by Condon et al, is close to 2.3, the standard value. Because of the poor resolution of IRAS, we do not know whether the bulk of the IRAS emission also comes from such compact regions. But, based on the brightness temperature and IRAS colours, Condon Broderick [8] conclude that most of the compact radio sources are optically thick at 25 μm. This is corroborated by the ground based high resolution mid-infrared observations of Soifer et al [12] that several of these compact sources are indeed very compact and contain a large fraction of the IRAS flux density. Based on radiation transfer calculation, one expects the 60 and 100 μm sizes to be not much larger. Hence, one may conclude that the radio-FIR relation also holds good for these compact cores. This raises another problem. Condon et al point out that for FIR luminosities in excess of $10^{11.5}$ $L_O$ the radiation density in a sphere of radius 100 pc is ~ 3 × $10^{-3}$ erg cm$^{-3}$; as a result, the inverse Compton scattering will become dominant and spoil the synchrotron-FIR correlation. For the synchrotron emission to dominate over the inverse Compton scattering, the magnetic field has to be larger than ~ $10^{-3}$ G. Such values are not unreasonable, if the expected magnetic field-matter density relation, $B \propto \rho^{0.5}$ holds good. Solomon et al [13] who observed a sample of ULIRGs in the CO line conclude that $\tau_{100\mu m}$, the FIR optical depth is high in these galaxies. Assuming an optical depth at 100 μm of unity and the radius of the sphere to be 100 pc and using the dust absorption coefficient given by Hildebrand [14] and a gas to dust ratio of 100, the average gas density is ~ 5000 cm$^{-3}$. Scaling from 5 μG at ρ ~ 1 cm$^{-3}$, we then obtain B ~ 350 μG for the compact source, about the same value as demanded for the dominance of synchrotron emission over the inverse Compton scattering. But these high values of B and ρ create other difficulties.

The synchrotron half life of an electron radiating at 10 GHz in a magnetic field of 350 μG is only 30,000 yrs. This is very much shorter than several million years, the period over which star formation and hence injection of electrons take place. We are then clearly in the 'calorimetric' regime in which the total energy of the electrons being continuously injected is converted into synchrotron radio emission. This helps to get a tight correlation. However, the equilibrium SED of electrons will steepen due to synchrotron loss leading to a steep radio index of -1.25 for the standard cosmic ray electron spectrum with an exponent of -2.5. This is contrary to the observational result of Niklas and Beck [15] who find that the distribution of spectral index for a large sample of galaxies peaks around -0.8 and less than 15% of the galaxies have a spectral index less than -1. But, the interesting question is, what is the spectral index of NT RC emission in ULIRGs?. What is usually measured is the total radio emission which is a mixture of thermal and NT emissions. Niklas et al [16] made accurate multi-band observations of a sample of 74 RSA galaxies and were able to separate the thermal and NT emissions. They find that at 1 and 10 GHz, the fraction of thermal emission is 10% and 30% respectively. Does the same hold good for ULIRGs? There are reasons to believe that the thermal fraction may be higher in ULIRGs.

The thermal fraction of about 30% at 10 GHz is derived for normal galaxies in which star formation is prevalent over 50 Myrs or more, greater than the life time of 8 $M_O$ stars, the lowest mass star that explodes as a type II SN which generate CR electrons. The ULIRGs could be much younger. If so, the NT radio emission will be less than the equilibrium value. Bressan et al [17] and Rengarajan & Mayya [18] have considered such scenarios and have computed the evolution of the FIR and NT radio emission. For continuous star formation, the FIR luminosity rises to an equilibrium value in 10-15 Myrs from the onset of star formation while the same takes more than 30-40 Myrs for the NT radio luminosity because the life time of 8 $M_O$ stars exploding as SN is that long.. But the thermal radio luminosity evolves on a time scale similar to or faster than the FIR time scale since it is mainly from living high mass stars. Thus, if the galaxy is observed within 10-20 Myrs of the onset of star formation, one expects to see a deficit in the NT emission, but not much in the FIR and thermal radio luminosity, thus leading to higher values of q and a larger fraction of thermal emission. The frequency of such sources will be higher if the duration of intense starbursts is not large, say, less than 30 Myrs. If the fraction of thermal emission is 2-3 times larger than in normal galaxies, we will observe a flat total spectrum even in the presence of a steep NT spectrum. In order to test the above scenario, we have to determine the NT spectral index; for this, we need highly accurate observations at several frequencies (5 or more) ranged over 1-50 GHz. Presently, we do not have such observations available. Some recent observations at 22 GHz by Prouton et al [19] reveal that a few ULIRGs having a q value larger than the mean by a factor of 2 or more, have rather flat 8.4-22 GHz spectra indicating the increase of thermal fraction. But, we need to await more accurate and more extensive high frequency observations to settle this issue. In conclusion, to avoid the adverse interference of inverse Compton scattering with the synchrotron-FIR correlation, either the injected electron spectrum has to be much flatter than the standard spectrum or the observed NT spectrum should be steep ( $\alpha \sim 1$ or more) accompanied by an increase in both the FIR-radio ratio and the thermal to NT ratio. Another test would be the observation of inverse Compton scattered high energy photons. Calculations have been performed for starburst galaxies ( Blom et al [20]), but they do not exist for the high density and high magnetic field regions discussed above. If B is indeed ~ 300 µG, the inverse Compton scattered photons will be in the 200 keV range, where sensitive measurements are not available. It will be of interest to perform the calculations and set limits on the magnetic field and radiation density in ULIRGs.

## 3. Role of Secondary Electrons

Niklas [21] noted an interesting feature of radio-FIR correlation . He analyzed the variation of the slope of the correlation with FIR luminosity. With sub samples of galaxies above different thresholds, he found an increase in the slope especially beyond log ($L_{FIR}$ /W) = 36.0. Most probably, the effect seen is associated with the NT emission. Niklas concludes that the increase of slope with luminosity is an intrinsic property of NT emission. Here, I suggest another reason for this.

In all the above discussions it has been assumed that the electron luminosity per SN event is the same in all galaxies and the luminosity density is proportional to the SFR. This is true only if the nature of the injected electrons is preserved during propagation in the ISM. For example, if re-acceleration takes place the luminosity density would increase and the spectrum may be flatter. At present, we do not know whether re-acceleration occurs and if it does, the parameters on which it depends. There is another way by which luminosity of electrons can increase. During their propagation in the ISM, the nucleonic component interacts with the atoms of the ISM and produce secondary electrons which are mostly ( ~ 80 %) positrons. The flux of secondary electrons depends on the total amount of matter traversed by the cosmic rays, i.e., the product of average gas density and the residence time. For the Galactic cosmic rays which typically traverse 3-4 g $cm^{-2}$ of matter, several calculations have been undertaken. We cite the results from the work of Moskalenko & Strong [22] who reproduce reasonably well the observed flux of positrons. They find that the secondary fraction is 0.24, 0.15 and 0.06 at 0.2, 1 and 5 GeV respectively. Thus for normal galaxies, the effect is small. However, the gas density in the sample of galaxies studied for the radio-FIR correlation range over a factor of more than 10. The Milky Way is roughly in the middle of the range. For galaxies that have a much higher density, the secondary contribution becomes significant. For e.g., if it is 5 times higher, the secondaries equal the primaries in the GeV range and the spectral index of the sum of primary and secondary electrons will increase by 15%. Since the magnetic field also increases with gas density, the GHz radiation will come more and more from ~ 1 GeV electrons. For e.g., the peak frequency for a Gev electron in a 100 µG field is close to a GHz. It should further be noted that due to the large range of frequencies over which the synchrotron radiation of a mono energetic electron is spread and given the steep powerlaw electron spectrum, 0.2-2 GeV electrons will significantly contribute to the GHz radio emission. As a result, for active and more luminous galaxies we may expect an increase by factor of 2 or more, in the NT emission and hence a similar decrease in the value of q. If an increase of a factor of 2 occurs over a decade in luminosity, the increase in the NT radio-FIR slope will be 0.3. This is quite close to the behaviour observed by Niklas (1997) for the more luminous galaxies.

For ULIRGs, both the magnetic field and the matter density are even higher in the compact cores. The matter density can be, as discussed in Sec.2, as high as 5000 $cm^{-3}$. For such high densities the amount of matter traversed will be several tens of g $cm^{-2}$ even for a life time of 10000 years. Since nucleons, unlike electrons, do not lose energy by synchrotron emission, they can propagate for a much longer time than electrons in these high magnetic field regions. Thus the matter traversed can be high even for lower densities like 1000 $cm^{-3}$ and the production of secondaries can exceed that of primary electrons. This can result in the lowering of the value of q, the FIR to radio ratio. Further, for magnetic fields of a few hundred µG, the GHz radiation is mainly from sub GeV electrons. The secondary electron spectrum is much flatter at these energies than at higher energies. Such a flat injected spectrum could then lead to a synchrotron loss steepened radio spectral index of -0.6 to -0.8 consistent with present observations. Also, the increased production of secondary electrons may compensate the expected reduction of SN generated primary electrons in the early stage of star burst discussed earlier.

If the material traversed is in excess of 50 g cm$^{-2}$, significant compared to the nucleon-nucleon interaction mean free path and the pair production mean free path for electrons, both the primary and secondary electron spectrum will be subject to brehmstrahlung losses and copious production of gamma rays will result from nuclear interactions. It is, therefore, of great interest to carry out such calculations and compare with observations in order to test such models or to set limits on the parameters involved.

## 4. Conclusion

The continuation of the radio-FIR correlation in ULIRGS which have compact cores of high photon density has been investigated. High magnetic field scaling with high density may be responsible in preserving the dominance of synchrotron radiation loss over the inverse Compton scattering process. However, this should result in a steep radio spectral index that is not observed. It is suggested that at high densities, the secondary electrons may dominate over the primary electrons. For high magnetic field of about $10^{-3}$ gauss, GHz radiation will be mostly from sub GeV electrons where the secondary spectrum is much flatter leading to a synchrotron loss steepened radio spectral index of only -0.8 to -1 consistent with observations. It is also pointed out that the FIR to radio ratio would be less than the equilibrium value for very young ( < 20 Myrs) star bursts. This ratio may further be affected by the increasing contribution from secondary electrons. To get a better insight into these problems, it is necessary to undertake accurate multi-frequency radio observations for a large sample of star burst galaxies and ULIRGs and attempt to separate the thermal and NT emissions and determine the NT spectral index.

## References

[1] J. M. Dickey and E. E. Salpeter, ApJ, 284, 461 (1985).
[2] G, Helou et al, ApJ, 298L, 7 (1905).
[3] T. de Jong, A&A, 147, L6 (1985).
[4] E. Wunderlich et al, A&AS, 69, 487 (1987)
[5] E. Hummel et al, A&A, 199, 91 (1988).
[6] A. Dressler, ApJ 236, 351 (1980).
[7] J. J.Condon and J. J. Broderick, AJ, 92, 94 (1986).
[8] J. J. Condon et al, ApJ, 376, 95 (1991).
[9] T. N. Rengaraja and K. V. K. Iyengar, MNRAS, 242, 74 (1990).
[10] U. Klein, A&A 246, 323 (1991).
[11] M. D. Bicay et al, ApJS, 98, 369 (1995).
[12] B. T. Soifer et al, AJ, 119, 509 (2000).
[13] P. M. Solomon et al, ApJ, 478, 144 (1997).
[14] R. H. Hildebrand, QJRAS, 24, 267 (1983).
[15] S. Niklas, and R. Beck, A&A, 320, 54 (1997).
[16] S. Niklas, A&A, 322, 19 (1997).
[17] A. Bressan et al, A&A, 392, 377 (2002).
[18] T. N. Rengarajan and Y. D. Mayya, AdSpR, 34, 675 (2004)
[19] O. R. Prouton et al, A&A, 421, 115 (2004).
[20] J. J. Blom et al, ApJ, 516, 744 (1999).
[21] S. Niklas, A&A, 322, 29 (1997).
[22] I. V. Moskalenko and A.W. Strong, ApJ, 493, 694 (1998.)